\documentclass{aa}  

\usepackage{epstopdf}
\usepackage{graphicx}
\usepackage{txfonts}
\usepackage{rotating}
\usepackage{subfigure}
\usepackage{gensymb}
\usepackage{tablefootnote}
\usepackage{graphicx}
\usepackage{subfig}
\usepackage[labelfont=bf]{caption}
\usepackage{natbib}
\usepackage{xcolor}

\bibpunct{(}{)}{;}{a}{}{,} 
\newcommand{\ipone}{IGR\,J17195-4100}

\newcommand{\nustar}{\textit{NuSTAR}}
\newcommand{\xmm}{\textit{XMM-Newton}}

\begin{document}

   \title{Significant reflection and absorption effects in the X-ray emission of the Intermediate Polar \ipone\ }
    
   \author{Elif \c{S}afak
          \inst{1,2}
          \and
          \c{S}ölen Balman\inst{2,3}
          \and
          Gloria Sala\inst{1,4}
          }

   \institute{Departament de Física, EEBE, Universitat Politècnica de Catalunya, c/Eduard Maristany 16, 08019 Barcelona, Spain
         \and
             Department of Astronomy and Space Sciences Faculty of Science, Istanbul University Beyazit, 34119, Istanbul, Turkey
             \and
              Faculty of Engineering and Natural Sciences, Kadir Has University, Cibali, Istanbul, 34083, Turkey
              \and
              Institut d’Estudis Espacials de Catalunya, c/Gran Capità 2–4, Ed. Nexus-201, 08034 Barcelona, Spain
             }

   \date{Received ...; accepted ...}

 
  \abstract
   {X-ray emission is emitted from shock-heated plasma in magnetic cataclysmic variables, particularly in Intermediate Polars, and is processed by absorption and scattering before reaching the observer.}
   {We investigate these effects in the X-ray emission of the Intermediate Polar \ipone\ by carrying out the X-ray spectral analysis, spin modulation and hardness ratio.}
  {We present high-sensitivity broadband X-ray spectral analysis by combining \nustar\ and \xmm\ observations in the 0.3-78.0 keV energy range. The X-ray spectral analysis is performed using six composite models, including the angle-dependent reflection model ({\tt reflect}), multi-temperature plasma emission models (CEVMKL, MKCFLOW), and photoionised and/or neutral partially covering absorption models ({\tt zxipcf, pcfabs}) within XSPEC. We examine also the spin modulation in four energy ranges and the hardness ratio, to determine the absorption and scattering effects.}
   {We find that the spectrum is best modelled with a reflection amplitude ($\Omega$) of 0.58 $^{+0.38}_{-0.26}$, an ionisation parameter, log($\xi$), of 1.46$^{+0.44}_{-0.23}$ with an equivalent hydrogen column density of 3.09$^{+2.26}_{-0.68}$ $\times$ 10$^{22}$ cm$^{-2}$, a neutral absorber, and a multi-temperature plasma temperature of 27.14$^{+2.0}_{-2.13}$ keV. In addition, we detect effects of electron scattering in the \nustar\ band, leading to modulation amplitude of about a steady 9\% that increases up to 15\% after 20 keV.}
   {We stress that these effects significantly affect the X-ray emission of intermediate polars and should be considered to obtain a good representation of the intrinsic spectrum.}


   \maketitle
%

\section{Introduction}

Magnetic cataclysmic variables (MCVs) are binary systems in which a late-type main sequence star accretes onto a magnetized white dwarf (WD) \citep{Warner1995}. Intermediate Polars (IPs), which is a subclass of the MCVs, have magnetic field strength in the range of 10$^5$ to 10$^7$ Gauss, allowing an accretion disk around the WD. The matter on the accretion disk is channelled to the magnetic field lines of the WD near the Alfven radius and falls freely through the gravitational potential towards the poles of the WD \citep{Patterson1994}.

Close to the surface of the WD, the infalling gas accelerates to high supersonic speeds, and then a strong steady shock forms above the poles of the WD, as it must decelerate.  In the shock, the kinetic energy of the infalling matter turns into thermal energy, resulting in the hot plasma that emits strong X-rays via optically thin thermal emission in the post-shock \citep{Ezuka_and_Ishida1999}. The shocked plasma has a temperature distribution that ranges from a maximum temperature at the shock gradually cooling through thermal bremsstrahlung until it settles on the WD surface \citep{Fabian1976}. The observed X-ray spectrum shows features such as $\sim$ 1 keV lines from the iron L shell and the presence of H- and He-like iron K-$\alpha$ emission lines at 6.9 keV and 6.7 keV, indicating the nature of optically thin thermal plasma emission \citep{Done_and_Magdziarz1998, Mukai2017, Hoshi1973}.

Assuming the X-ray source emits isotropically as expected, some fraction of the X-rays can be reflected by the WD surface and also reprocessed by relatively cold regions, such as the pre-shock region, the WD surface, etc. \citep{Done_and_Osborne_and_Beardmore1995, Lightman_and_White1998}. Thus, the X-ray spectrum of IPs can be described not only by the nature of the multi-temperature plasma but also by the inclusion of scattering and absorption properties.

The X-rays can impinge on the relatively cold white dwarf surface and then be reflected by cold electrons, which includes Compton processes. The reflection from a relatively cold, optically thick surface, such as the WD surface, causes the Fe K-$\alpha$ fluorescent line emission at 6.4 keV and the Compton reflection hump that hardens the spectrum in the energy range 10-30 keV \citep{George_and_Fabian1991, vanTeeseling_etal1996, Beardmore_etal1995, Magdziarz_and_Zdziarski1995, Done_and_Magdziarz1998, Ezuka_and_Ishida1999}. These effects of reflection in the X-ray spectrum have been previously studied in MCVs \citep{Rothschild_etal1981, Beardmore_etal1995, vanTeeseling_etal1996, Cropper_etal1998, Revnivtsev_etal2004, Beardmore_etal2000, Mukai_etal_2015, Dutta_and_Rana2022, Hayashi_etal2011, Hayashi_etal2021, Joshi_etal2023}. Given that reflection modifies the intrinsic spectrum, it is important to consider the reflection component, especially in X-ray spectral analyses to estimate physical quantities of IP systems, such as WD mass, radius, and height of the shock \citep{Hayashi_etal2011}. Previous research has shown that the reflected spectrum is angle-dependent becoming harder as the viewing angle, cos\textit{i}, increases (angle \textit{i} measured with respect to the normal of the reflecting surface) \citep{Hua_and_Lingenfelter1992, Magdziarz_and_Zdziarski1995}. However, not only the shape of the reflected spectrum depends on the viewing angle, but also many factors influence the reflection effect. A small shock height increases the reflection effect as it enhances the solid angle of the WD surface with respect to the accretion column, but for the same geometry the equivalent width of the Fe K-$\alpha$ fluorescence line is decreased \citep{Luna_etal2018, Hayashi_etal2018, Beardmore_etal2000}.
Similarly, massive WD and therefore a larger specific accretion rate increases the reflection effect because it results in a hotter and shorter shock height \citep[cf. Fig. 22][]{Hayashi_etal2018}. Also, the equivalent width of the Fe K-$\alpha$ fluorescent line is angle-dependent and decreases with increasing \textit{i} \citep{George_and_Fabian1991, Hayashi_etal2018},  whereas it increases with increasing iron abundance \citep{Hayashi_etal2018}.

IPs are known to show complex absorber features in soft X-rays. The X-ray emission region is covered by absorbing material, some fraction of the X-ray photons have to travel through this region before they reach the observer \citep{Done_and_Osborne_and_Beardmore1995, Balman2012}. This intrinsic absorption originates from relatively cool regions of the system, such as the surface of the WD, accretion column, accretion curtain, accretion disk, etc. The absorber material cannot be characterized only by neutral matter, because the hard X-rays can partially ionise the absorber material \citep{Cropper_etal1998, Dutta_and_Rana2022, Islam_and_Mukai2021, Pekon2012}. Thus, especially below 4.0 keV, the neutral and/or ionised absorption features can be seen in soft X-rays. Similar to Compton reflection, absorption causes spectral hardening, but in relatively softer X-ray spectra \citep{Done_and_Magdziarz1998}.

The absorption, occultation, and scattering can lead to modulations of the X-ray light curve due to the spin or orbital period. The X-ray spin and orbital modulations have been investigated in many MCVs and IP systems by \citet{Hellier1993, Parker2005, Norton1989, Norton1997, Kim1995, Evans2004, 2024Balman}. The accretion flow above the shock causes attenuation of the X-ray emission through absorption and electron scattering. Additionally, the WD surface can occult the X-ray source, reducing the X-ray emission. In each case, the X-ray emission decreases periodically due to WD spin, causing modulation of the spin-folded X-ray light curves. In soft X-rays, photoelectric absorption dominates spin modulation due to its energy dependence and is expected to decrease with increasing energy. However, the spin modulation depends also on the geometry and the presence of other absorption mechanisms within the system \citep{Norton1989, Kim1995}.

\ipone\ was discovered by \citet{Bird2004} in the \textit{INTEGRAL} Galactic Plane survey conducted in February and October 2003. \citet{Barlow2006} then carried out hard X-ray spectral analysis in the 20-100 keV energy range using \textit{INTEGRAL}/IBIS data and obtained a temperature of 27.0 $\pm$ 4.4 keV with the X-ray flux of 2.46 $\times$ 10$^{-11}$ erg s$^{-1}$ cm$^{-2}$ as the best-fitted result. Through investigation of the optical spectrum obtained from observations of the \textit{CTIO} telescope, \citet{Masetti2006} identified the source as a MCV. Afterwards, \citet{Butters2008} classified \ipone\ as Intermediate Polar and determined its orbital period to be approximately 1.7 hours using the \textit{RXTE} observation.  \citet{Pretorius2009} further calculated the orbital period of 4.005 $\pm$ 0.006 h and spin period of 1139.55 $\pm$ 0.03 s with optical time-resolved observations. \citet{Bernardini_etal2012} also measured the spin period of 1062 $\pm$ 2 s using the \xmm\ observations. Then, \citet{Girish_and_Singh1992} calculated the new period of the \ipone\ using \xmm\ and \textit{Suzaku} observations with a spin period of 1053.7 $\pm$ 12.2 s and an orbital period of 3.52 $^{+1.43}_{-0.80}$ h. The mass estimate of \ipone\ has been conducted by several studies. \citet{Yuasa_etal2010} estimated 1.03$^{-+1.24}_{-0.22}$ M$_{\odot}$ with a temperature of 59.6 keV using \textit{Suzaku} in the 3.0 - 50 keV energy range, and \citet{Suleimanov2019} found 0.72 $\pm$ 0.06 M$_{\odot}$ using \nustar\ and \textit{Swift}/BAT data above 20 keV, and lastly \citet{Shawn2020} obtained 0.84$^{+0.8}_{-0.7}$ M$_{\odot}$ along with a temperature of 22.9$^{+3.6}_{-2.9}$ keV using \nustar\ observations in the 20-78 keV energy range.

In this study, we investigate the broadband spectral analysis of  \ipone\ using \xmm\ and \nustar\ observations in the range of 0.3-78.0 keV with six composite models, examining the presence of reflection and ionised and/or neutral absorption effects with different multi-temperature plasma models. Unlike previous studies, in the present work \nustar\ data were used, including the whole 3.5-78.0 keV energy range. It is also the first time that the ionised absorption and angle-dependent reflection effects are considered for this source and investigated in the 0.3-78 keV range for  \ipone. In addition, we present spin modulation analysis in four different \nustar\ energy ranges to detect the modulations due to possible absorption and electron scattering effects.

\renewcommand{\arraystretch}{1.5}
\begin{table*} 
    \centering
    
    \caption{Description of six composite models}
    \begin{tabular}{cc} 
    \hline 
    \hline
    \multicolumn{1}{c}{Model Name}  &
    \multicolumn{1}{c}{The Composite Models}   \\
    
     \hline
     M1 & \texttt{cons$\times$phabs$\times$pcfabs$\times$(reflect$\times$mkcflow+gauss+gauss)} \\  
     \hline

     M2 & \texttt{cons$\times$phabs$\times$pcfabs$\times$zxipcf$\times$(reflect$\times$mkcflow+gauss+gauss)} \\
     \hline

     M3 & \texttt{cons$\times$phabs$\times$pcfabs$\times$(reflect$\times$cevmkl+gauss+gauss)} \\  

     \hline

     M4 & \texttt{cons$\times$phabs$\times$pcfabs$\times$zxipcf$\times$(reflect$\times$cevmkl+gauss+gauss)} \\
     \hline

     M5 & \texttt{cons$\times$phabs$\times$zxipcf$\times$zxipcf$\times$(reflect$\times$mkcflow+gauss+gauss)} \\  

     \hline

     M6 & \texttt{cons$\times$phabs$\times$zxipcf$\times$zxipcf$\times$(reflect$\times$cevmkl+gauss+gauss)} \\
     \hline
     
    \end{tabular}\\    
    \label{table:Table1}
\end{table*}

\section{Observation and data reduction}\label{sec:obs}

In this study, we use the publicly available archival  X-ray data from the  \nustar\ and \xmm Observatories.  \ipone\ was observed by  \nustar\ (OBSID=30460005002) for a duration of 29.5 ks on October 25, 2018 and by \xmm\ (OBSID=0601270201) for  a duration of 33.9 ks on September 3, 2009 published in \citet{Bernardini_etal2012}.

 \nustar\, launched by NASA in 2012, is the first X-ray telescope in orbit to image in the 3 - 79 keV band. It is designed with two similar Wolter I-type X-ray optics (OMA and OMB) focussed on two similar focal plane modules (FPMA and FPMB) \citep{Harrison_etal2013, Madsen_etal2015}. We reprocessed the  \nustar\ data using  \nustar\ Data Analysis Software (NuSTARDAS v2.1.2), part of HEASOFT v6.30, with standardized filtering criteria and CALDB (20220331) files to obtain high-quality science data. The main NuSTARDAS script {\tt nupipeline} was used to generate calibrated and cleaned science-quality event files. We then, selected the source and background circular regions using DS9 and extracted the 70\arcsec\ and 100\arcsec\ regions from the clean event files, respectively (the size of the background regions were normalised to the source regions). Finally, we used the {\tt nuproduct} task to generate the light curves, sky images, and energy spectra using the source and background regions, and grouped all spectra with at least 40 counts per bin using the {\tt grppha}  FTOOLS command. 

\xmm\ is launched in 1999 by the European Space Agency (ESA), observing the X-ray, ultraviolet, and optical sky with the six coaligned instruments, including three EPIC X-ray imaging cameras (PN, MOS1, MOS2) and two X-ray RGSs (Reflection Grating Spectrometers) and the OM (Optical Monitor) \citep{Jansen_etal2001,Lumb_etal2012}. The \xmm\ energy range of 0.1-10 keV is an important contribution to spectral analysis in the soft X-rays. We have used the EPIC pn data due to its high sensitivity and high time resolution. The EPIC pn data of the source were reduced and calibrated by SAS (Scientific Analysis System) software (v19.1.0) with the latest calibration files using {\tt epicproc} task. 

We then, checked a light curve in the range of 10 keV to 12 keV from the event file to detect possible solar flares. Following this inspection, we created good time intervals to filter out the detected solar flares from the data. Since the source is bright, we checked the generated event file for pile-up with the {\tt epatplot} task. In order to remove the identified pile-up effect, we excluded the circular source region  of size 6\arcsec\ from the bright central point. Consequently, the source photons are extracted from an annular region of inner radius 6\arcsec\ and outer radius 40\arcsec\ from the source position, and the background photons are extracted from a circular region of 40\arcsec. Even after the pile-up removal, we restricted our analysis of EPIC pn data to energies lower than 8 keV to avoid the slight residual spectral hardening observed in the {\tt epatplot} results after pile-up mitigation.

Finally, we used the task {\tt evselect} to generate the source and background spectra using defined regions from the filtered event file and including photons with a quality flag equal to zero and a pattern less than or equal to four. The redistribution matrix and ancillary file were created using SAS tasks, RMFGEN and ARFGEN, respectively. The spectra were grouped with the {\tt grppha} task using a minimum of 150 counts per bin. 

For further joint spectral analyses using two different missions, HEASOFT v6.30 \footnote{https://heasarc.gsfc.nasa.gov/} and XSPEC v12.14 \footnote{https://heasarc.gsfc.nasa.gov/xanadu/xspec} are utilized.

In order to perform spin modulation analysis, we chose only FPMA and FPMB data of \nustar, as \citet{Bernardini_etal2012} had previously studied the spin modulation of \ipone\ using \xmm. The background-subtracted light curves of \nustar\ data are generated using {\tt nuproduct} task of NuSTARDAS. Utilizing the {\tt pilow} and {\tt pihigh} parameters of the {\tt nuproduct} task, we produced these light curves in different energy ranges using the channel-energy conversion equation\footnote{E = Channel Number * 0.04 keV + 1.6 keV} for \nustar. The spin period of \ipone\ is found to be 1057.3 s by  \citet{Girish_and_Singh1992} using the \xmm\ and {\it Suzaku} missions. We employed the XRONOS software package tool {\tt efold} to fold light curves at the spin period of 1057.3,  and produced spin pulsed-profile and hardness ratios.

\section{Analysis}
\renewcommand{\arraystretch}{1.9}
\begin{table*}

\centering
    \caption{The best-fit spectral parameters of the source \ipone\ with errors at 90\% confidence level.}
    
    \begin{tabular}{lccccccc} 
    \hline 
    \hline
    \multicolumn{1}{l}{Parameter}  &
    \multicolumn{1}{c}{Unit}  &
    \multicolumn{1}{c}{M1}  &
    \multicolumn{1}{c}{M2}  &
    \multicolumn{1}{c}{M3}  &
    \multicolumn{1}{c}{M4}  &
    \multicolumn{1}{c}{M5}  &
    \multicolumn{1}{c}{M6}    \\  
    
    \hline
    N$_{\rm{H}}$$^{phabs}$ & 10$^{22}$cm$^{-2}$ & 0.11$^{+0.004}_{-0.008}$ & 0.12$^{+0.01}_{-0.005}$ &  0.11$^{+0.005}_{-0.005}$ & 0.12$^{+0.007}_{-0.003}$ & 0.13$^{+0.006}_{-0.005}$ & 0.12$^{+0.01}_{-0.007}$\\
     
    N$_{\rm{H}}$$^{pcf}$ & 10$^{22}$cm$^{-2}$ & 3.26$^{+0.23}_{-0.48}$ & 14.51$^{+6.68}_{-2.03}$ & 4.79$^{+0.53}_{-0.52}$ & 11.82$^{+3.9}_{-1.8}$ & N/A & N/A \\

    \emph{cvf}$^{pcf}$ & & 0.55$^{+0.01}_{-0.01}$ & 0.42$^{+0.02}_{-0.12}$ & 0.53$^{+0.01}_{-0.01}$& 0.38$^{+0.02}_{-0.15}$& N/A & N/A \\
    
    N$_{\rm{H}}$$^{zxipcf}$ & 10$^{22}$cm$^{-2}$ & N/A & 3.33$^{+1.9}_{-0.7}$ & N/A & 3.09$^{+2.26}_{-0.68}$ & 3.0$^{+0.1}_{-0.82}$ & 2.95$^{+1.82}_{-1.19}$\\
    
     &  & & & & & 7.36$^{+1.53}_{-0.93}$& 7.39$^{+1.64}_{-1.09}$\\

    \emph{cvf}$^{zxipcf}$ & &N/A & 0.54$^{+0.04}_{-0.07}$ & N/A & 0.44$^{+0.09}_{-0.04}$ & 0.49$^{+0.06}_{-0.04}$ & 0.39$^{+0.08}_{-0.03}$\\
    
    &  & & & & & 0.46$^{+0.04}_{-0.12}$ & 0.43$^{+0.12}_{-0.06}$\\
    
     \emph{$\log\xi$$^{1}$} & & N/A & 1.47$^{+0.22}_{-0.17}$ & N/A & 1.46$^{+0.44}_{-0.23}$ & 1.63$^{+0.34}_{-0.22}$ & 1.69$^{+0.4}_{-0.31}$\\

    &  & & & & & -0.18$^{+0.3}_{-0.25}$ & 0.17$^{+0.92}_{-0.32}$\\
     
    $\Omega$$^{2}$ & & 13.04$^{+5.7}_{-2.27}$& 0.49$^{+1.48}_{-0.21}$& 2.11 $^{+1.16}_{-0.59}$ & 0.58 $^{+0.38}_{-0.26}$ & 0.69$^{+0.29}_{-0.17}$ & 0.54$^{+0.2}_{-0.19}$\\
     
    \emph{cosi} & & $\le$ 0.05 & $\ge$ 0.16 & 0.19$^{+0.08}_{-0.08}$  & $\ge$ 0.36 & $\ge$ 0.64 & $\ge$ 0.51\\
     
    T$^{3}$$_{\rm{max}}$ & keV & 25.91$^{+2.26}_{-3.67}$ & 30.4$^{+4.37}_{-2.92}$ & 27.82$^{+1.97}_{-2.06}$ & 27.14$^{+2.0}_{-2.13}$ & 28.19$^{+1.78}_{-1.23}$ & 26.57$^{+1.59}_{-1.21}$ \\

    Z$^{4}$ & \(Z_\odot\) &0.21$^{+0.4}_{-0.4}$ & 0.27$^{+0.05}_{-0.04}$ & 0.3$^{+0.05}_{-0.05}$ & 0.36$^{+0.06}_{-0.04}$& 0.25$^{+0.02}_{-0.02}$ & 0.36$^{+0.05}_{-0.04}$\\
    
    Norm$^{5}_{\rm{mkc,cev}}$ & & 2.7$^{+0.2}_{-0.1}$ & 3.4$^{+0.2}_{-0.5}$  & 6.1$^{+0.01}_{-0.01}$ & 6.8$^{+0.2}_{-0.2}$ &  3.4$^{+0.1}_{-0.2}$ & 6.8$^{+0.02}_{-0.2}$ \\
    
    E$_{l,\rm{Fe K\alpha}}$ & keV & 6.4$^{+0.01}_{-0.03}$ &6.4$^{+0.01}_{-0.02}$  & 6.4$^{+0.01}_{-0.01}$ &6.4$^{+0.007}_{-0.02}$  & 6.4$^{+0.006}_{-0.02}$  & 6.4$^{+0.005}_{-0.01}$   \\

    Norm$^{6}_{\rm{Fe K\alpha}}$ & phot.\ cm$^{-2}$s$^{-1}$ & 6.2$^{+0.06}_{-0.06}$  & 6.8$^{+0.6}_{-0.6}$  & 6.6$^{+0.5}_{-0.5}$  & 6.5$^{+0.6}_{-0.8}$  & 6.4$^{+0.7}_{-0.7}$  & 6.6$^{+0.6}_{-0.7}$  \\
     
    E$_{l,\rm{OVII}}$ & keV & 0.57$^{+0.006}_{-0.006}$ & 0.57$^{+0.01}_{-0.01}$ &0.56$^{+0.007}_{-0.006}$ & 0.56$^{+0.007}_{-0.008}$ &0.57$^{+0.007}_{-0.008}$ & 0.56$^{+0.008}_{-0.008}$ \\
    
    Norm$^{6}_{\rm{OVII}}$ & phot.\ cm$^{-2}$s$^{-1}$  &0.9$^{+0.1}_{-0.1}$ &1.0$^{+0.2}_{-0.3}$ &0.9$^{+0.1}_{-0.2}$ &0.7$^{+0.2}_{-0.2}$ &1.0$^{+0.1}_{-0.1}$ & 0.7$^{+0.2}_{-0.1}$ \\ 
    
    $\chi^2_\nu / dof$ & & 1.28/1081 & 1.11/1078 & 1.19/1081 & 1.08/1078 & 1.09/1077 & 1.08/1077\\
    
    \hline
    \end{tabular}
    
    \tablefoot{
        \\$^{1}$ $\log\xi$ : The ionisation parameter
        \\$^{2}$ $\Omega$ : The amplitude of the reflection.
        \\$^{3}$ \textit{alpha} for CEVMKL  is set at 1.0 assuming the standard cooling flow model. 
        \\$^{4}$ Z: The plasma abundance of MKCFLOW model and the Fe abundance of CEVMKL model. The {\tt angr} table of abundances is used \citep{Anders_and_Grevesse989}.
        \\$^{5}$The norm of the MKCFLOW and CEVMKL models is in the units of 1$\times$10$^{-8}$ \(M_\odot\)/yr and 1$\times$10$^{-2}$N$_{cevmkl}$, respectively.
        \\$^{6}$The norm of GAUSS model for Fe K$\alpha$ and O VII lines are in units of 1 $\times$ 10$^{-5}$ and 1 $\times$ 10$^{-3}$  \emph{photons/cm$^2$/s}, respectively; photon flux in the line.
        \\The sigma ($\sigma$) of GAUSS models are fixed at 1.0 $\times$ 10$^{-3}$ assuming narrow lines limited with the spectral resolution. 
       \\Iron abundance and abundance parameter of the \texttt{reflect} model is kept at 1.0 Solar abundance.\\
         The X-ray Flux in the 3.5-10 keV energy range is 2.6 $\times$ 10$^{-11}$ erg s$^{-1}$ cm$^{-2}$  for the \xmm\ EPIC pn, and 2.9 $\times$ 10$^{-11}$ erg s$^{-1}$ cm$^{-2}$  and 2.8 $\times$  10$^{-11}$ erg s$^{-1}$ cm$^{-2}$  for FPMA and FPMB data of \nustar, respectively.
        \\The cross-normalisation constant is fixed to 1.0 for the FPMA yielding 0.98 and (0.91-0.94) for FPMB and pn data, respectively.

    }
    \label{table:Table2}
\end{table*}

\subsection{Spectral analysis }

The X-ray spectral analysis was performed in the 0.3-78 keV range using X-ray Spectral Analysis Software package XSPEC v12.12.1 \citep{Arnaud1996}. The combined spectra consist of \xmm\ EPIC pn data in the range of 0.3-8.0 keV and \nustar\ data in the range of 3.5- 78.0 keV. Pile-up is mitigated as described in Sec. \ref{sec:obs}. The X-ray Flux in the common energy range of 3.5-10.0 keV for the two instruments is obtained as 2.6 $\times$ 10$^{-11}$ erg s$^{-1}$ cm$^{-2}$ for EPIC pn, and 2.9 $\times$ 10$^{-11}$ erg s$^{-1}$ cm$^{-2}$  and 2.8 $\times$  10$^{-11}$ erg s$^{-1}$ cm$^{-2}$  for FPMA/B, respectively.

\begin{figure*}[ht!]
\hspace{-0.2cm}
        \includegraphics[width=.53\linewidth]{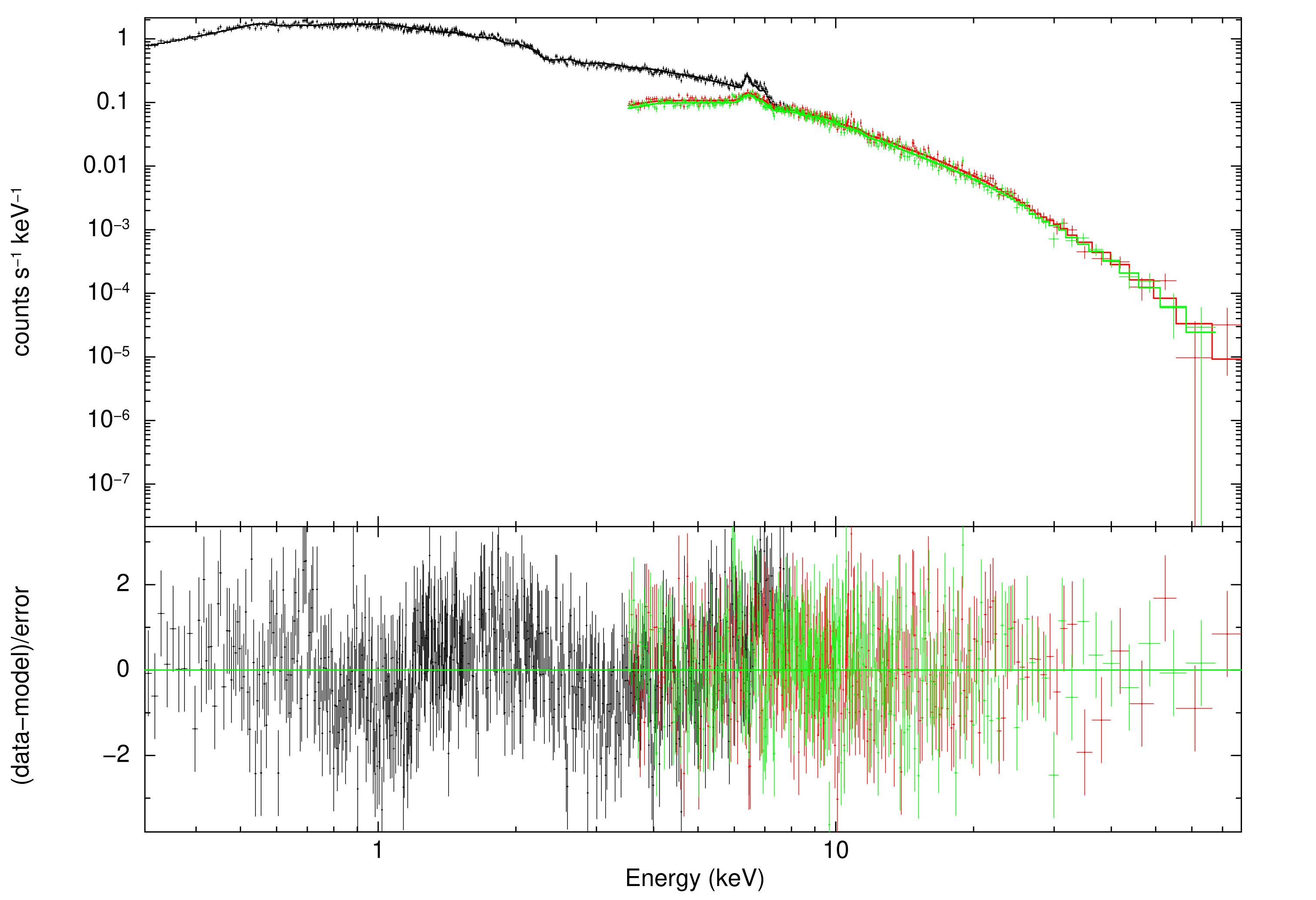}
    \hfill
    \hspace{-0.6cm}
        \includegraphics[width=.53\linewidth]{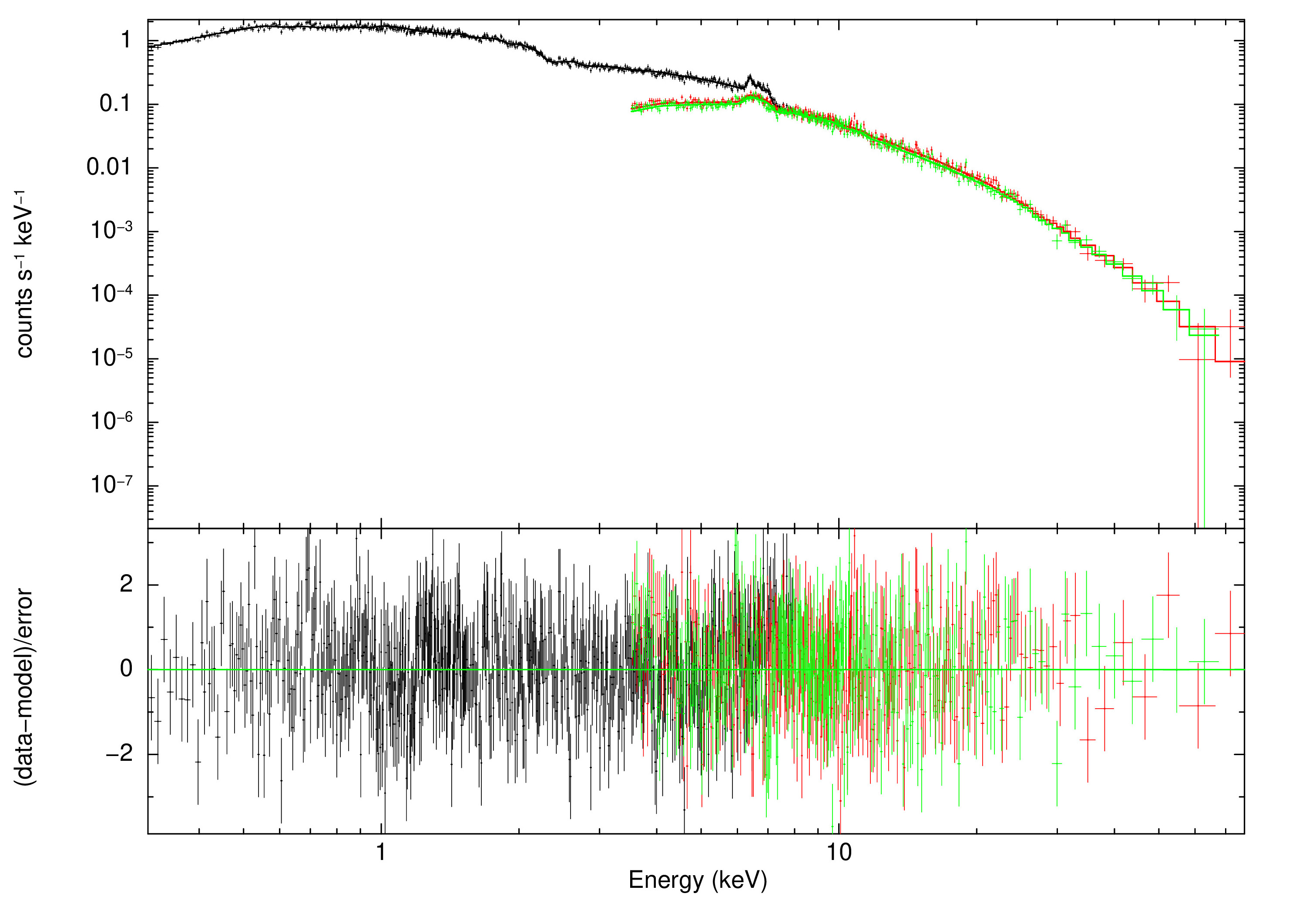}
    \hfill
    \caption{The fitted X-ray spectrum of \ipone\ obtained from \xmm\ (black), \nustar/FPMA(red) and \nustar/FPMB (green) fit with the composite models M3 (\textit{left}) and  M4 (\textit{right}). The second panel shows the residuals in standard deviations.}
    \label{fig:fig1}
\end{figure*}

As emphasized earlier, the X-ray emission from magnetic CVs are modified by the reflection effect from the white dwarf surface which can cause the Compton Hump in the range of 10 - 30 keV, and the fluorescent Fe K${\alpha}$ emission line at 6.4 keV, together with the intrinsic absorption. For the analysis, we have constructed six composite models including two different plasma emission models which are listed in Table \ref{table:Table1}. The spectral parameters resulting from the fits using these six composite models are detailed in Table \ref{table:Table2}. In addition, two of the fitted composite models are displayed in Figure \ref{fig:fig1}.

A cross-normalisation constant (model) was included in all composite models to properly combine the three instruments in order to adjust their calibration discrepancies. This parameter is used by fixing FPMA to 1, leaving the others free. In addition, the photoelectric absorption due to the interstellar medium was fit using the \texttt{phabs} model. We also tested the \texttt{tbabs} model but did not proceed with it, as it failed to improve the fit beyond the \texttt{phabs} model. The column density, $N_{H}$, was iterated in the fitting process, starting from the value of 0.6 $\times$ 10$^{22}$ cm$^{-2}$ calculated from the HI4PI database \citep{HI4PICollaboration}.

Since X-ray photons encounter absorbing medium before reaching the observer, X-ray spectra include absorption effects. To characterize the continuum adequately, we have paid attention to the components of the absorbing media. Thus, we created three different scenarios to obtain an adequate description of the soft X-rays. The scenarios were constructed considering the absorbing material consisting of  i) only a neutral absorber (models M1 and M3), ii) neutral and ionised absorbers (models M2 and M4), and iii) two ionised absorbers with two different ionisation parameters (models M5 and M6). The neutral absorber is theoretically part of the accretion column \citep{Mukai2017}. However, the regions more likely close to the standing shock can also be ionised by high-energy photons. We, therefore, construct (ii) and (iii) scenarios. In (ii)  ionized absorber is one, in (iii) there are two ionized absorbers, one close to the emitting region (PSR), perhaps in the accretion curtain, and the other is further out in the disc likely close to the stream-disk impact zone. We have applied models assuming that the X-ray emission region is partially covered by absorbing regions. 

The photoelectric absorption effect was modelled with \texttt{pcfabs} model that assumes partially covering absorption by neutral material in the line of sight. There are two free parameters of the model, the column density of the absorber, $N_{H}$, and the covering fraction, \textit{cvf}. The absorption from the partially ionised gas was modelled by partially covering of a partially ionised absorber, \texttt{zxipcf} model \citep{Reeves2008}. This model consists of four parameters, equivalent column density of the partially ionised absorber, $N_{H}$, logarithm of the ionisation parameter, $\log\xi$, the covering fraction, \textit{cvf}, and one fixed parameter, redshift, which is fixed to zero. The ionisation parameter, $\xi$, is defined as $\xi  = L_{X}/nr^{2}$, where $L_{X}$ is the X-ray luminosity, $n$ is the number density in the absorbing gas, and $r$ is the distance between the X-ray source and absorbing material. As seen in Table \ref{table:Table1}, the models M1 and M3 include only \texttt{pcfabs}, M2 and M4 include \texttt{pcfabs$\times$zxipcf}, and models M5 and M6 include \texttt{zxipcf$\times$zxipcf}.

\begin{figure}[h]
    \centering
    \includegraphics[width=\linewidth]{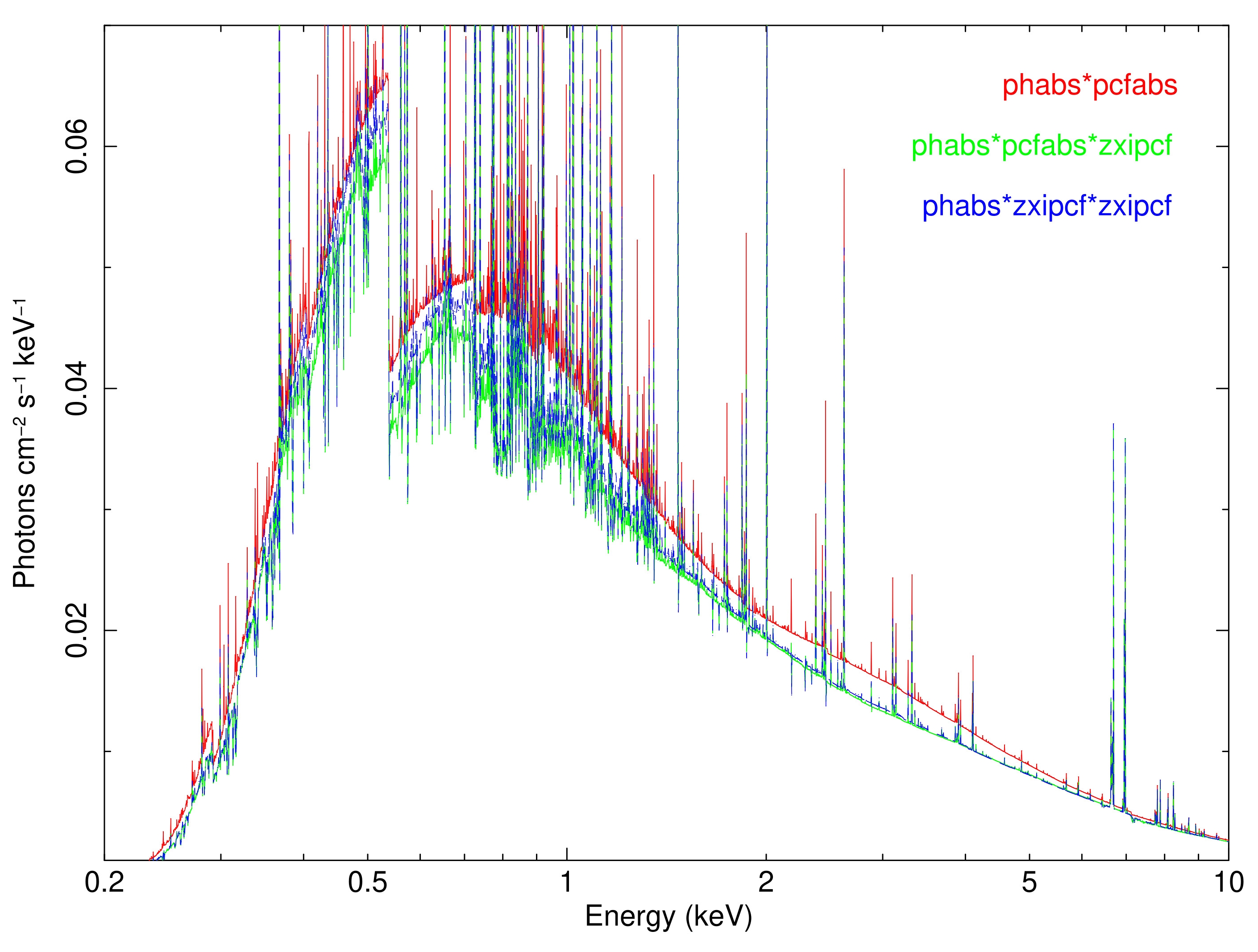}
    \caption{The fitted absorption components as in Table \ref{table:Table2}. Represented composite models are M3, M4 and M6.}
      \label{fig:fig2}
\end{figure}

Figure \ref{fig:fig2} shows the plots of three composite-fitted models (only) applied to the spectra as detailed in Table 2. It is clear that there are absorption differences in the models. The red-colored line shows the fitted model with { \tt pcfabs} (only neutral absorption), the green-colored line shows the one with the \texttt{pcfabs}$\times$\texttt{zxipcf} absorption component (one neutral and one ionised absorber) and the blue-colored line represents the fitted model with \texttt{zxipcf}$\times$\texttt{zxipcf} (two ionised absorbers). Especially, the ionised absorption model plays an important role, the green line shows several absorption lines/features additional to the model shown by the red line, and the blue line indicates similarly deeper features/lines. Moreover, evident from Figure 2, significant spectral continuum modification is seen between the red line and the blue or green lines in the energy range of 0.4-5.0 keV. Our fits do not differentiate between the two ionised absorber included models due to moderate spectral resolution. The model { \tt pcfabs} modifies the continuum via neutral absorption to a given level as seen in the red line in Figure 2, but the models with ionized absorbers make the fits significantly better showing more modification via absorption lines and continuum absorption.

The Compton scattering effect, our other main research in this study, has signatures in the X-ray spectrum with a hardening in the energy range 10$-$30 keV and a prominent emission at 6.4 keV, the fluorescent Fe K$\alpha$ line.  To model the reflected spectrum, we used the model by \citet{Magdziarz_and_Zdziarski1995}, {\tt reflect}, and included it in all composite models. This reflection model is called the angle-dependent reflection, which considers the reflection depending on the viewing angle of the reflector, in this case, the WD surface. \citet{Magdziarz_and_Zdziarski1995} show that the angle-dependent reflected spectrum is similar to the average angle reflected spectrum at the inclination $\mu$ = 0.45.  Hence, we used the inclination angle fixed to 0.45 at the start of the fitting and then varied the inclination angle, which improved the fit. The amplitude parameter of the model, the reflection factor ($\Omega$), was set free and starting from 1.0. The {\tt reflect} model also has three fixed parameters: redshift, the metal abundance, and the iron abundance, which were kept fixed in the fit. We used metal and iron abundances fixed at solar values for the reflection model. We checked that tying up the reflection component abundances to plasma abundances has no effect on our fit results.

We finally added a narrow emission line using the GAUSS model in order to account for the fluorescent iron K$\alpha$ reflection line. The sigma ($\sigma$) is fixed at 1 $\times$ 10$^{-3}$ keV, considering a narrow line limited by the resolution of the detectors. 

The collisionally ionised multi-temperature plasma in the post-shock region is the source of the X-rays and cools by bremsstrahlung emission. We use two different isobaric cooling flow type multi-temperature plasma models, MKCFLOW \citep{Mushotzky_and_Szymkowiak1988} and CEVMKL \citep{Singh1996} in the different composite models to model the X-ray emission. We set the low temperature to 0.0808 keV,  the minimum value allowed by the model, and the redshift parameter to 1.0 $\times$ 10$^{-7}$ (for the distance of 639.4 pc obtained from \citet{Bailer2021}), in MKCFLOW model. In the CEVMKL model, we use the power law index of temperature distribution, $\alpha$ parameter, fixed at 1 (for a collisionally ionised typical cooling flow model). For CEVMKL model, when we fit all abundances to the solar values, the plasma temperature is overestimated and thus the intrinsic spectrum is not well-represented. Instead, we set only Fe abundance free and it improved the fit considerably. In both plasma models, we chose the switch as 2 that uses the ATOMDB \footnote{http://www.atomdb.org/} database. The abundance table is selected as 'angr' \citep{Anders_and_Grevesse989} using abund command. As follows from Table \ref{table:Table1}, MKCFLOW model is included in the composite models M1, M2 and M5, while CEVMKL model is included in the M3, M4 and M6 composite-models.

In addition, the X-ray spectrum of  \ipone\ exhibits an emission line at 0.57 keV corresponding to the O VII line (He-like Oxygen line). The ionised emission line was previously found by \citet{Bernardini_etal2012}. Thus, we have added a GAUSS model to all composite models to account for this line. The sigma of this emission line is set to 1$\times$10$^{-3}$ keV as found in the study of \citet{Bernardini_etal2012}.

The best-fit results of the X-ray analyses performed using the six composite models are presented in Table \ref{table:Table2}, with error ranges at the 90\% confidence level for a single parameter. It can be seen that the model improves when we consider that the intrinsic absorbing material consists of neutral and ionised gas. The absorbing region close to the standing shock can be ionised by hard X-rays. The findings show that hard X-rays from the shock region ionise the absorber material in the system. We also obtained more consistent reflection amplitude values ($\Omega$ $<$ 1) in the fits when the ionised absorber model was included. We found that the ionised equivalent hydrogen column density was $N_{H}$ $\sim$ 3.09$\times$ 10$^{22}$ cm$^{-2}$, and the coverage of the ionised material was $\sim$ 46\%, and the ionisation parameter of the material was $\sim$ 1.56.  In the scenario where we consider absorbing ionised gas with  two different ionisation parameters (with no neutral intrinsic absorber), the fits do not improve. This shows that the neutral absorber is significant in the X-ray spectra. 

To evaluate the significance of each absorption components, we performed an F-statistic, defined as the ratio of the reduced chi-squared values \citep{Bevington2003,Orlandini2012,Iyer2015}. The F-statistic gives a confidence level of 99.1\% (2.6 $\sigma$) and 95\% (1.98 $\sigma$) between models M1 and M2 and models M3 and M4, respectively. This result indicates that the effects of the ionised absorbers are significant and such effects are present in the X-ray spectra. We also compared models M5 and M6, the models with two ionised absorption models, with models M2 and M4 and obtained confidence levels of 53\% (0.73 $\sigma$) and 61.7\% (0.87 $\sigma$), respectively. Therefore, we can not significantly attribute if there are two different warm absorbing regions and existence of a single ionized absorber or two such ionized absorbers are equally likely since all such fits have good reduced chi-squared values around 1.1. An intrinsic neutral absorption is necessary for the X-ray spectrum, which is theoretically required for IPs \citep{Mukai2017}.

Both plasma models describe the joint-spectra well and estimate the plasma temperature nearly the same, along with the low solar abundance of metals or iron. The reflection amplitude($\Omega$) is $\sim$ 0.57 with an inclination angle, \textit{cosi}, $\ge$ 0.16 when the ionised absorber model was included in the composite models. We carried out another test of the significance for the reflection effect in the X-ray spectrum using the method of \citet{Mukai_etal_2015}, when we set the reflection amplitude to 0.0 in the M2 and M4 models. We used the F-test in XSPEC to compare the goodness of the fits with the cases where the {\tt reflect} model was included, we observed that the plasma temperature increased to 42.7 keV and 34.7 keV (when {\tt reflect} was not used), and the F-test yielded a probability of 2 $\times$ 10$^{-5}$ and 7 $\times$ 10$^{-5}$, respectively, which showed that reflection significantly (over 3$\sigma$) affects X-ray spectra and hardens the spectrum, causing an overestimation of the plasma temperature, if neglected.

\subsection{Spin modulation analysis}

The attenuation of X-ray flux due to absorption and/or scattering leads to modulation at the spin period. We used the \nustar\ data to examine the changing of X-ray flux as a result of scattering and absorption at the spin phase,  and generated spin-folded X-ray light curves using the period, 1057.3 s. In addition, different energy ranges were selected to investigate the energy dependence due to absorption or scattering. We chose four specific energy ranges as 3$-$5 keV, 5$-$10 keV, 10$-$20 keV, and 20$-$50 keV to generate spin-folded light curves. The modulation in the different energy bands were calculated using pulsed-fraction defined as $P= (F_{max}-F_{min})/(F_{min}+F_{max})$, where F denotes the X-ray flux. We obtained minimum and maximum flux values by fitting a sine model to the spin-folded light curves. 

\begin{figure}[h!tbp]
\hspace{-0.5cm}
        \includegraphics[height=8.2cm,width=\linewidth]{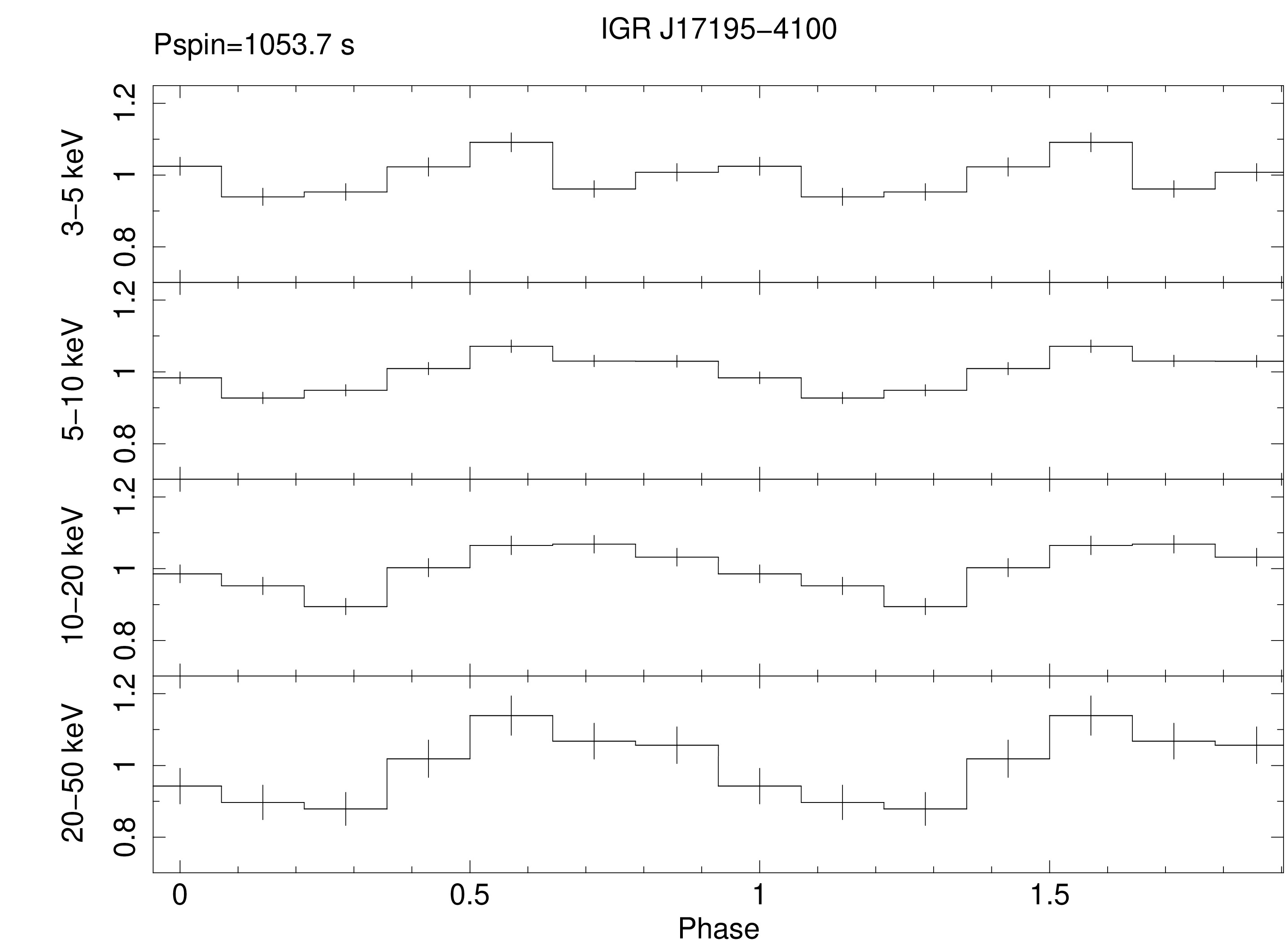}
        \caption{The normalised (to average count rate) spin X-ray pulse profile of \ipone\ using NuSTAR Telescope. The phase zero is arbitrary.}
          \label{fig:fig3}
\end{figure}

The spin-pulsed profiles are displayed in Fig. \ref{fig:fig3}. The spin phase zero is arbitrary in these profiles. We observed that the spin modulation does not diminish in the \nustar\ energy bands with nearly a constant value (see Table \ref{table:Table3}). The calculated pulsed fractions in Table \ref{table:Table3} show that the modulation is slightly higher in the 20-50 keV energy range relative to the lower energy ranges. \citet{Bernardini_etal2012} found modulation amplitude below 3.0 keV, 14$\pm$1\% in 0.3-1.0 keV energy range and 13$\pm$1\% in 1.0-3.0 keV energy range, which is too low for the photoelectric effect to dominate and more consistent with electron scattering effects \citep{Rosen1992}. Also, since photoelectric absorption is energy-dependent, $E^{-3}$, we expect the modulation to decrease steeply as the energy increases. However, the spin modulation of \ipone\ does not decrease in the higher energy ranges but flattens out, pointing to electron scattering. Electron scattering can cause non-isotropic hard X-rays depending also on geometry, leading to spin modulation in the harder X-rays, especially above 6 keV \citep{Rosen1992, Buckley1989}. Therefore, we assume that X-rays are scattered by electrons residing in the ionised gas in the system. This gas is most likely associated with the ionised absorber we detect in the source.

As a further inspection, we performed the hardness ratio (HR) using 10$-$20/3$-$5 keV and 20$-$50/5$-$10 keV energy ranges. The hardness ratio compares the number of detected photons in two energy ranges. The analysis provides an adequate comparison of the X-ray variability with respect to the spin phase in different energy bands. We have carried out this analysis using FPMA and FPMB data. Figure \ref{fig:fig4} shows the hardness ratios comparing the low-energy band to the high-energy band. These results again imply the presence of scattering which is directionally dependent.  As displayed in Table \ref{table:Table3}, the pulsed fraction has a value around (7-9)\% consistent with electron scattering which in general shows low levels of expected modulation amplitudes in the X-rays \citep{Rosen1992}. We find that this pulsed fraction increases out to 15\% in the hardest X-ray band (20-50 keV). We interpret this as an effect of  Compton hump (existing Comptonization)  as a result of reflection expected in the same energy band. The increase of hardness seen in the spin phase 0.5 corresponds to the phase where the maximum X-ray flux is reached. We assume
that the accretion column points away from the observer yielding
a clear view of the X-ray emitting post-shock region. However, our interpretation requires the determination of system parameters such as the orbital inclination in order to be verified.

We do not consider the self-occultation effect for spin modulation, which is the occultation of X-ray emission regions by the white dwarf as it rotates, leading to spin modulation \citep{Evans2005MNRAS}. The spin pulse profile shows only one X-ray maximum in a phase, indicating a single pole is visible to the observer, and it does not have a large modulation depth revealing that the inclination angle is low.

\section{Discussion}

The main motivations for our research are to investigate the reflection and absorption effects on X-ray emission. For this purpose, we have conducted broadband X-ray spectrum analysis and spin pulsed profile and hardness ratio studies.

The X-ray broadband spectra were carried out in the 0.3-78 keV energy range using highly sensitive observations of \ipone\, \nustar\ and \xmm. In order to truly characterize the X-ray spectrum, several phenomena are taken into account through this research. Our main considerations were to study the Compton reflection effect on X-ray spectra and to determine the neutral and/or ionised nature of the absorbers. The comprehensive spectral analysis is performed by combining FPMA and FPMB data of \nustar\ in the 3.5-78.0 keV range and EPIC pn data of \xmm\ in the 0.3-8.0 keV energy range. 

We have constructed six composite models to provide a good description of the broadband X-ray spectra of \ipone. The composite models were created considering different absorption components, neutral and/or ionised, and different plasma models, MKCFLOW and CEVMKL (in XSPEC). Table \ref{table:Table2} shows the best-fit results of six composite models with errors at 90\% confidence level. The spectral results of CEVMKL and MKCFLOW plasma models are compatible within the uncertainties. We achieved the best fit with reduced chi-square of 1.11/1078 and 1.08/1078 in the M2 and M4 composite models, respectively. This leads to the conclusion that broadband X-ray spectra are well fit with a combination of partially covering neutral and ionised absorber models. 

\begin{figure*}[ht!bp]
\hspace{-0.5cm}
            \includegraphics[height=8.2cm,width=.51\linewidth]{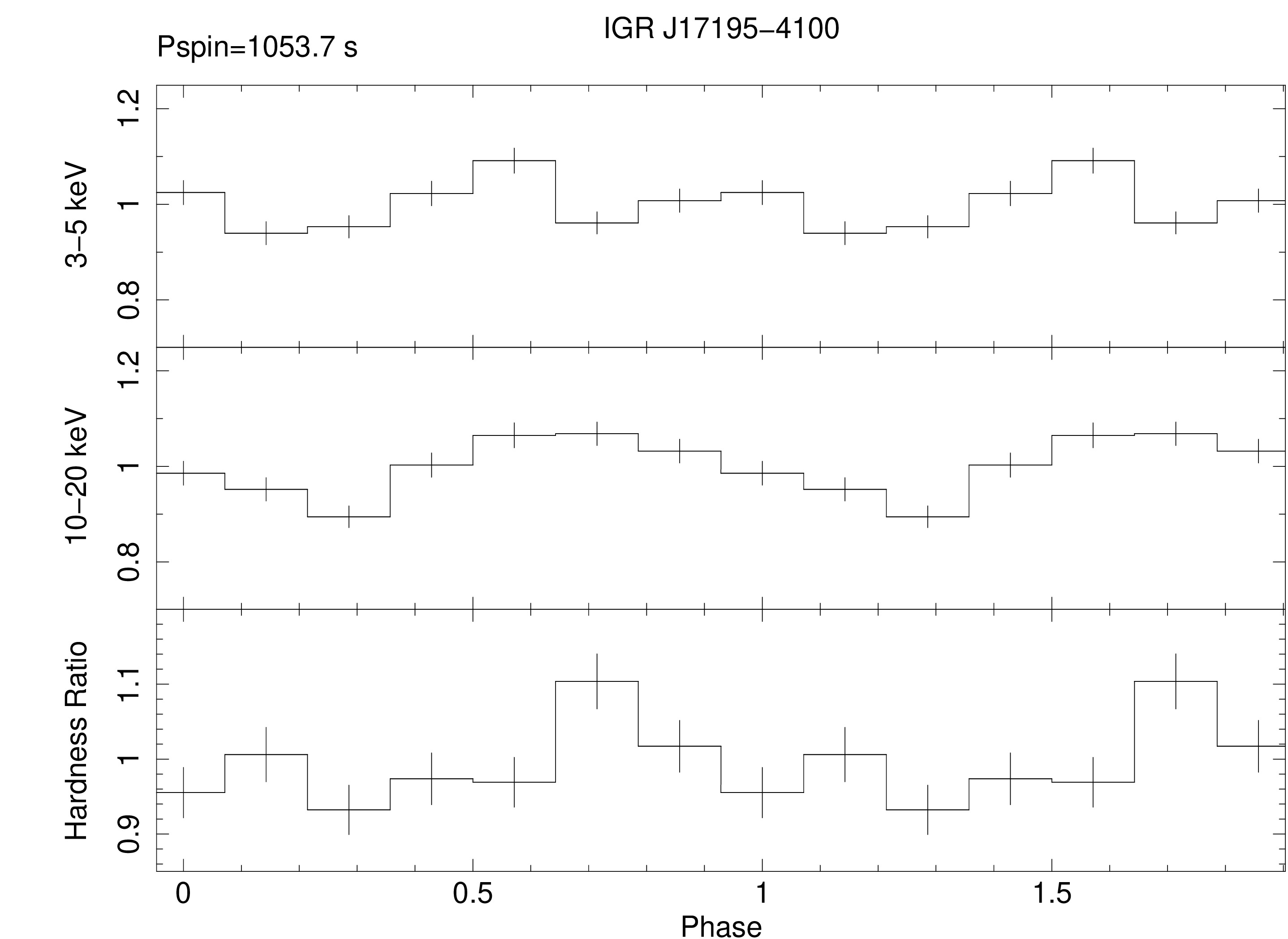}
        \hfill
        \hspace{-0.5cm}
            \includegraphics[height=8.2cm,width=.51\linewidth]{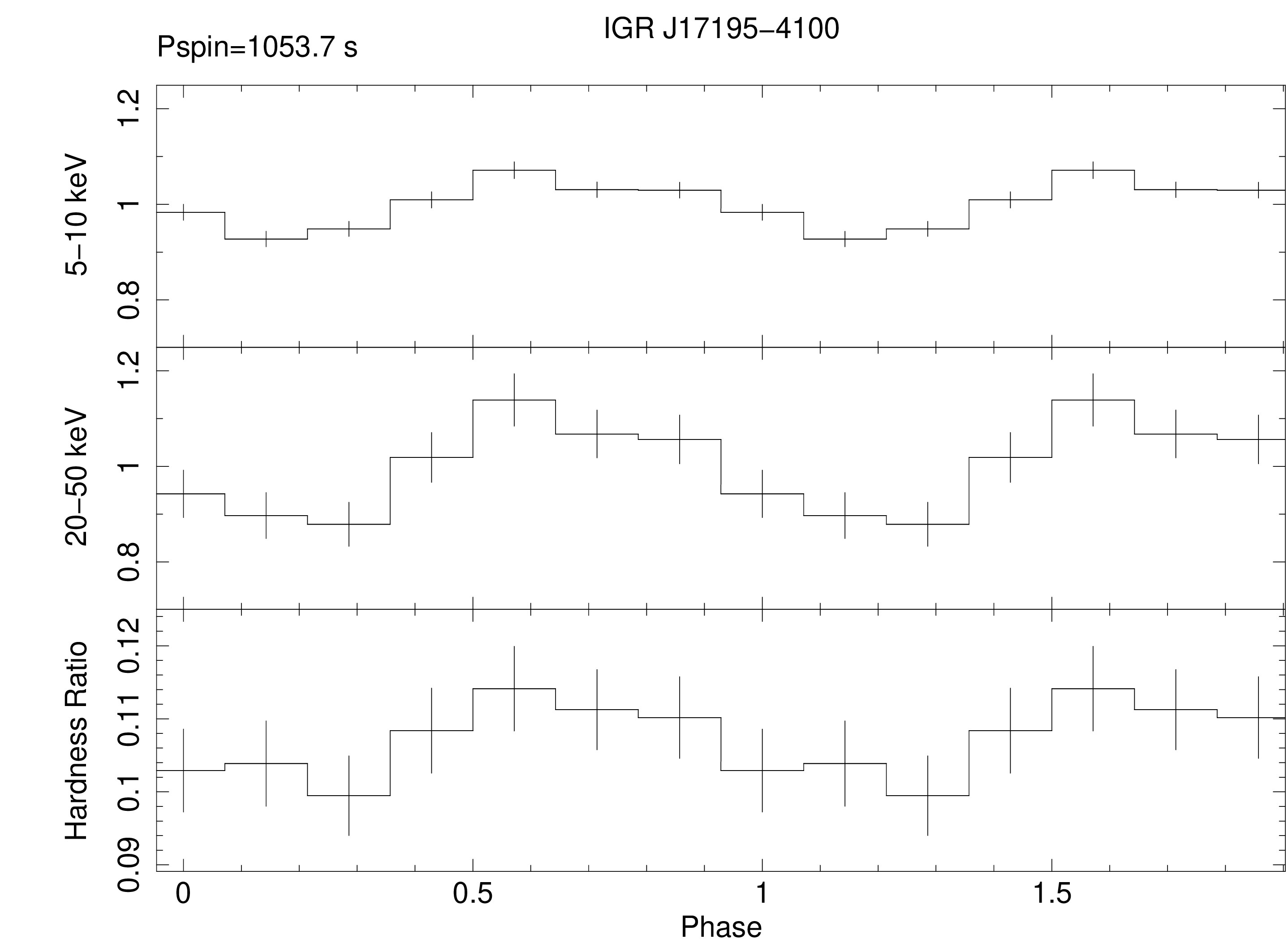}
        \caption{The hardness ratio in the X-ray light curve folded at the spin period of \ipone\ using \nustar\ Telescope. The phase zero is arbitrary.}  \label{fig:fig4}
\end{figure*}

We obtain a plasma temperature compatible within errors with \citet{Shawn2020} study. Since \citet{Shawn2020} eliminates absorption and reflection effects using energy >20 keV, the approximate plasma temperature result obtained from the broadband X-ray analysis indicates that the composite models we used represent reflection and absorption effects well. Thus, accounting for both the reflection and ionised absorption, allowed us to obtain the true plasma temperature.

We found the hydrogen column density and covering fraction of the neutral absorber model to be 14.51$\times$ 10$^{22}$ cm$^{-2}$ and 11.82$\times$ 10$^{22}$ cm$^{-2}$ with 42\% and 38\% covering fractions using the models M2 and M4, respectively. In contrast to previous studies, the neutral absorber values we measured are higher  \citep{Girish_and_Singh1992, Bernardini_etal2012, Butters2008}. On the other hand, using M1 and M3 models, which include only a neutral absorber model,  slightly lower neutral absorber values (less by a factor of 2) are calculated which are less than values found in the previous studies, except for \citet{Butters2008} study, which found a neutral absorber of 0.7$\times$10$^{22}$ cm$^{-2}$ in the 2.5-20 keV energy range.

Although ionised absorbers have been the subject of several investigations for IPs \citep{Islam_and_Mukai2021, Dutta_and_Rana2022, Joshi_etal2023, Pekon2012}, it was considered for the first time for spectral analysis of the \ipone. The equivalent hydrogen column density and covering fraction were found to be (2.6-5.4)$\times$ 10$^{22}$ cm$^{-2}$, with  (44-54)\% covering fraction. The ionisation parameter was found to be almost the same, log($\xi$)=1.47 and 1.46, in models M2 and M4, respectively. As we highlighted earlier, accounting for the presence of ionised absorber in the system improved the fits and provided an essential contribution to the X-ray spectral analysis.

The effect of Compton reflection on the X-ray emission in  IP systems has been considered by several studies \citep{vanTeeseling_etal1996, Beardmore_etal1995, Done_and_Osborne_and_Beardmore1995, Hayashi_etal2011, Hayashi_etal2021, Mukai_etal_2015, Cropper_etal1998, Revnivtsev_etal2004, Hayashi_etal2018, Beardmore_etal2000}. The hard X-ray photons can be downscattered by relatively cold electrons, leading to the Compton Hump in the energy range 10-30 keV. In this research, we investigated the significance of the reflection in the X-ray emission of  \ipone\ for the first time by inserting an angle-dependent reflection model, reflect, in all composite models. We obtained larger reflection amplitudes using the models M1 and M3, including only the neutral absorber. After adding the ionised absorber model to our composite models, we achieved a reflection amplitude smaller than 1. We suggest that the ionised absorber model fits the spectra well and provides more consistent values for the reflection amplitude in the spectral analysis. The F-test results performed on the models  M2 and M4 reveal that the reflection has a significance of more than 3$\sigma$ in the X-ray emission, leading to an overestimation of the plasma temperature of  \ipone\  by about 30\% if the reflection effect is ignored. Therefore, taking into account the reflection effect on X-ray emission is necessary to accurately characterize the system spectral parameters.

The Galactic column density along the line of sight of \ipone\ is 0.6 $\times$ 10$^{22}$ cm$^{-2}$, estimated using \citet{HI4PICollaboration}. As shown in  Table \ref{table:Table2}, the absorption from the interstellar medium was found lower than expected. Previous analyses also obtained a slightly lower absorption for interstellar medium \citep{Tomsick_etal2006, Bernardini_etal2012, Girish_and_Singh1992}. Since the low N${_H}$ value fits this soft region well, we did not include additional plasma models such as a MEKAL model to account for the slight soft excess.

An ionised oxygen line, O VII, observed at 0.57 keV has been detected by \citet{Bernardini_etal2012, Maiolino2021} studies. We have also detected this O VII emission line at 0.57 keV in this study and modelled it with a GAUSS model. This emission line can be interpreted as another evidence of the presence of ionised material in the system. In addition, \citet{Bernardini_etal2012} detected three more emission lines, O VIII, Ne IX, and Ne X, observed at 0.65 keV, 0.92 keV and 1.02 keV, respectively. We did not include these lines as they were not found to be significant in our spectral analysis.We note that the RGS spectra (1,2) are indicative of residual variations that may be of absorption features but no significant result can be obtained since the sensitivity difference between EPIC pn and RGS is about 0.01-0.005 considering the RGS energy range which degrades flux effectively.

The unabsorbed X-ray luminosity in the energy range 0.3-78.0 keV is calculated as 5.2$\times$10$^{33}$ erg s$^{-1}$ using the XSPEC task {\it flux} and a distance of 621.8 pc \citep{Bailer2021}. We also calculated a mass accretion rate of 3.1 $\times$ 10$^{16}$ g s$^{-1}$ using WD radius of 6.7$\times$10$^{8}$ cm from mass-radius relationship \citep{Nauenberg1972}, assuming WD mass of 0.84 M$_{\odot}$ \citep{Shawn2020}. Furthermore, we calculated the ionised absorber distance from the X-ray source using the number density of 10$^{15}$ cm$^{-3}$ \citep{Mukai2017} and a log($\xi$) of 1.46 from the results of model M4. We obtained this distance to be 4$\times$10$^{8}$ cm, pointing to a region close to the WD, likely the accretion curtain.

The spin pulse profile and hardness ratios are shown in Fig. \ref{fig:fig3} and Fig. \ref{fig:fig4}, respectively. The energy bands in the hard X-rays still shows modulations at the spin period with almost a constant modulation percentage that increases slightly between 20-50 keV range. The modulation in the soft  X-ray bands are expected to occur due to photoelectric absorption, which is energy dependent. As the energy increases, the modulation caused by photoelectric absorption is expected to gradually decrease. But in the \nustar\ bands above 3 keV and even above 10 keV, we still observe modulation, as seen in Table \ref{table:Table3}. We attribute this low percentage modulation in the hard bands to electron scattering \citep{Rosen1992, Buckley1989}. The X-ray photons can be scattered from free electrons, causing modulation due to the change in the direction of the X-ray photons. We observe the pulse peaks in more or less the same spin phase as the modulation in the most bands so we see the same geometry. We believe that the maximum X-ray flux is reached when the pole points away from the observer, and that when the pole points towards the observer, the X-ray photons are absorbed and scattered by the accretion flow, resulting in modulation in the same spin phase. Furthermore, in case of small $\Delta$R/R, scattering occurs close to the WD surface \citep[see Fig. 4 of][]{Rosen1992}, resulting in both effects being observed in the same phase. We note that we have not conducted a phase-resolved spectroscopy of the source because of the low modulation amplitude of ~ 10\% almost over the entire band of 0.3-78.0 keV which may not allow for a good statistical separation of the spectra at different phases properly. 

\citet{Hayashi_etal2018} clearly shows that the reflected spectrum depends on the mass of the WD, the abundances, the specific accretion rate, as well as the inclination angle. Assuming the distance of the ionised material (absorber) from the X-ray source as an estimate for the height of the accretion column, 4$\times$10$^{8}$ cm, and with $R_{WD}$ calculated as 6.7$\times$10$^{8}$ cm, we can assume that the WD has a tall accretion column. In the scenario of a tall accretion column, our findings are consistent with the \citet{Hayashi_etal2018} study. Since the tall accretion column reduces the solid angle of the WD, it can reduce the Compton hump, especially at large viewing angles \citep[see Fig. 6 of ][]{Hayashi_etal2018}. This could explain the low reflection amplitude, $\Omega$ $<$ 1, obtained in the composite models M2, M4, M5 and M6, which are the best-fit results we achieved (see Table \ref{table:Table2}).

\section{Conclusions}

\renewcommand{\arraystretch}{2.5}
\begin{table*}[ht!bp]
    \centering
    \caption{The Pulse Fraction (\%) of X-ray spin modulation of \ipone\ in four different energy ranges.}\label{tab3}
    \begin{tabular}{lcccc} 
    \hline 
    \hline
    \multicolumn{1}{l}{Telescope}  &
    \multicolumn{4}{c}{Pulse Fraction \% } \\
     & 3.0-5.0 keV & 5.0-10.0 keV & 10-20 keV & 20-50 keV\\
     \hline
     \hline
     NuSTAR&8.9$\pm$3 & 8$\pm$1 & 8.4$\pm$2 & 12$\pm$3 \\
     \hline     
    \end{tabular}
    \label{table:Table3}
\end{table*}

MCVs and IPs comprise more than 30\% of CV population and they are known, particularly IPs, to be the hardest X-ray emitters and perhaps the brightest for this matter. MCVs are readily detected in X-ray surveys (XMM-Newton, Swift, INTEGRAL) and studied since they have luminosities several times 10$^{(30-34)}$ erg/s contributing to the Galactic X-ray luminosity function; also, playing a crucial role in understanding Galactic X-ray binary populations  \citep[see][for the surveys]{deMartinoetal2020,2020Lutovinov}. As a result, understanding the spectral characteristics and their hard X-ray emission becomes essential, in general, to be able to classify and study this class of emitters.  Therefore, we have undertaken a previously known hard X-ray IP, \ipone, and investigated the hard X-ray spectrum in a broad range (0.3-78.0 keV) using joint \xmm\ and \nustar\ data to see how reflection and ionised absorption affected the spectrum of this source. We find that reflection from the surface of the WD in the system is consistent with the spectral characteristics at more than 3$\sigma$ confidence level. We also observe that models including both partially ionised and neutral absorber models better represent the soft X-ray spectra, indicating that ionised absorbers (i.e., warm absorbing regions) are  present in the system along with neutral absorbing gas. Both of these characteristics harden the broadband X-ray spectrum. These also indicate that Compton scattering and electron scattering are influential on the broadband X-ray spectrum.

The investigation of the ionised and neutral absorbers along with the reflection effect on the X-ray spectra help us to predict the system mechanism. The ionisation parameter, log($\xi$), derived from the best-fit results indicates that ionised material is present along with neutral material in the accretion column. The calculated distance of the ionised absorbing material led us to consider the tall shock scenario.

In this tall accretion column scenario where the WD will subtend a smaller solid angle, we measure a relatively smaller reflection amplitude, $\Omega$ $<$ 1, together with a cos$i$ $>$ 0.2, consistent with the previous theoretical calculations on reflection effect and its spectrum. We also found that the low metal abundances we obtain are consistent with the expected equivalent width of the Fe K$\alpha$ fluorescent line source detected by \citet{Bernardini_etal2012}. Nevertheless, the underlying origin of the low metal abundance and low iron abundance of \ipone\ observed in both plasma models and in previous studies is not well understood.

The spin modulation analysis reveals that the X-rays are under the effect of low photoelectric absorption and electron scattering with a modulation amplitude about (8-14)\% over the 0.3-78.0 energy band. Especially the modulation observed at high energies can be explained by electron scattering, which can change the direction of the X-ray photons as reported in \citet{Rosen1992} study.

We conclude that both reflection and ionised absorber effects characterize the X-ray spectrum of \ipone\ with high significance, and should be considered when analysing X-ray emission to obtain the intrinsic source emissions in the X-rays.

\begin{acknowledgements}
We would like to thank the anonymous referee for useful comments and valuable suggestions that improved the quality of the manuscript. This paper is jointly led by E\c{S} and \c{S}B which formed a major part of the MS thesis of E\c{S} (supervisor \c{S}B), accepted by Istanbul University, Dept. of Astronomy and Space Sciences and Institute of Sciences. E\c{S} and \c{S}B acknowledge  support by the Scientific Research Projects Coordination Unit of Istanbul University through the BAP Project No: 40017. E\c{S} acknowledges funding by l’Agència de Gestió d’Ajuts Universitaris i de Recerca (AGAUR) Generalitat de Catalunya official predoctoral program under the FI-SDUR grant (2023-FISDU-00225). E\c{S} and GS acknowledge support by the Spanish MINECO grant PID2023-148661NB-I00, by the E.U. FEDER funds, and by the AGAUR/Generalitat de Catalunya grant SGR-386/2021. This research is based on observations obtained with \xmm, and \nustar. The former is an ESA science mission with instruments and contributions directly funded by ESA Member States and the National Aeronautics and Space Administration, NASA and latter is a mission led by the California Institute of Technology, managed by the Jet Propulsion Laboratory, and funded by NASA.

\end{acknowledgements}

%
%
\bibpunct{(}{)}{;}{a}{}{,} 
\bibliographystyle{aa}
\bibliography{refl_abs}{}

\begin{thebibliography}{67}
\expandafter\ifx\csname natexlab\endcsname\relax\def\natexlab#1{#1}\fi

\bibitem[{{Anders} \& {Grevesse}(1989)}]{Anders_and_Grevesse989}
{Anders}, E. \& {Grevesse}, N. 1989, \gca, 53, 197

\bibitem[{{Arnaud}(1996)}]{Arnaud1996}
{Arnaud}, K.~A. 1996, in Astronomical Society of the Pacific Conference Series, Vol. 101, Astronomical Data Analysis Software and Systems V, ed. G.~H. {Jacoby} \& J.~{Barnes}, 17

\bibitem[{{Bailer-Jones} {et~al.}(2021){Bailer-Jones}, {Rybizki}, {Fouesneau}, {Demleitner}, \& {Andrae}}]{Bailer2021}
{Bailer-Jones}, C.~A.~L., {Rybizki}, J., {Fouesneau}, M., {Demleitner}, M., \& {Andrae}, R. 2021, \aj, 161, 147

\bibitem[{{Balman}(2012)}]{Balman2012}
{Balman}, S. 2012, \memsai, 83, 585

\bibitem[{{Balman} {et~al.}(2024){Balman}, {Khamitov}, {Kolbin}, {{\c{C}}al{\i}{\c{s}}kan}, {Bikmaev}, {{\"O}zd{\"o}nmez}, {Burenin}, {K{\i}l{\i}{\c{c}}}, {Eseno{\u{g}}lu}, {Yelkenci}, {{\c{C}}amurdan}, {Gilfanov}, {Nas{\i}ro{\u{g}}lu}, {Sonba{\c{s}}}, {Gabdeev}, {Irtuganov}, {Sayga{\c{c}}}, {Nikolaeva}, {Sakhibullin}, {Er}, {Sazonov}, {Medvedev}, {G{\"u}ver}, \& {Fi{\c{s}}ek}}]{2024Balman}
{Balman}, {\c{S}}., {Khamitov}, I., {Kolbin}, A., {et~al.} 2024, \aap, 684, A190

\bibitem[{{Barlow} {et~al.}(2006){Barlow}, {Knigge}, {Bird}, {J Dean}, {Clark}, {Hill}, {Molina}, \& {Sguera}}]{Barlow2006}
{Barlow}, E.~J., {Knigge}, C., {Bird}, A.~J., {et~al.} 2006, \mnras, 372, 224

\bibitem[{{Beardmore} {et~al.}(1995){Beardmore}, {Done}, {Osborne}, \& {Ishida}}]{Beardmore_etal1995}
{Beardmore}, A.~P., {Done}, C., {Osborne}, J.~P., \& {Ishida}, M. 1995, \mnras, 272, 749

\bibitem[{{Beardmore} {et~al.}(2000){Beardmore}, {Osborne}, \& {Hellier}}]{Beardmore_etal2000}
{Beardmore}, A.~P., {Osborne}, J.~P., \& {Hellier}, C. 2000, \mnras, 315, 307

\bibitem[{{Bernardini} {et~al.}(2012){Bernardini}, {de Martino}, {Falanga}, {Mukai}, {Matt}, {Bonnet-Bidaud}, {Masetti}, \& {Mouchet}}]{Bernardini_etal2012}
{Bernardini}, F., {de Martino}, D., {Falanga}, M., {et~al.} 2012, \aap, 542, A22

\bibitem[{{Bevington} \& {Robinson}(2003)}]{Bevington2003}
{Bevington}, P.~R. \& {Robinson}, D.~K. 2003, {Data reduction and error analysis for the physical sciences}

\bibitem[{{Bird} {et~al.}(2004){Bird}, {Barlow}, {Bassani}, {Bazzano}, {Bodaghee}, {Capitanio}, {Cocchi}, {Del Santo}, {Dean}, {Hill}, {Lebrun}, {Malaguti}, {Malizia}, {Much}, {Shaw}, {Stephen}, {Terrier}, {Ubertini}, \& {Walter}}]{Bird2004}
{Bird}, A.~J., {Barlow}, E.~J., {Bassani}, L., {et~al.} 2004, \apjl, 607, L33

\bibitem[{{Buckley} \& {Tuohy}(1989)}]{Buckley1989}
{Buckley}, D.~A.~H. \& {Tuohy}, I.~R. 1989, \apj, 344, 376

\bibitem[{{Butters} {et~al.}(2008){Butters}, {Norton}, {Hakala}, {Mukai}, \& {Barlow}}]{Butters2008}
{Butters}, O.~W., {Norton}, A.~J., {Hakala}, P., {Mukai}, K., \& {Barlow}, E.~J. 2008, \aap, 487, 271

\bibitem[{{Cropper} {et~al.}(1998){Cropper}, {Ramsay}, \& {Wu}}]{Cropper_etal1998}
{Cropper}, M., {Ramsay}, G., \& {Wu}, K. 1998, \mnras, 293, 222

\bibitem[{{de Martino} {et~al.}(2020){de Martino}, {Bernardini}, {Mukai}, {Falanga}, \& {Masetti}}]{deMartinoetal2020}
{de Martino}, D., {Bernardini}, F., {Mukai}, K., {Falanga}, M., \& {Masetti}, N. 2020, Advances in Space Research, 66, 1209

\bibitem[{{Done} \& {Magdziarz}(1998)}]{Done_and_Magdziarz1998}
{Done}, C. \& {Magdziarz}, P. 1998, \mnras, 298, 737

\bibitem[{{Done} {et~al.}(1995){Done}, {Osborne}, \& {Beardmore}}]{Done_and_Osborne_and_Beardmore1995}
{Done}, C., {Osborne}, J.~P., \& {Beardmore}, A.~P. 1995, \mnras, 276, 483

\bibitem[{{Dutta} \& {Rana}(2022)}]{Dutta_and_Rana2022}
{Dutta}, A. \& {Rana}, V. 2022, \apj, 940, 100

\bibitem[{{Evans} \& {Hellier}(2004)}]{Evans2004}
{Evans}, P.~A. \& {Hellier}, C. 2004, \mnras, 353, 447

\bibitem[{{Evans} \& {Hellier}(2005)}]{Evans2005MNRAS}
{Evans}, P.~A. \& {Hellier}, C. 2005, \mnras, 359, 1531

\bibitem[{{Ezuka} \& {Ishida}(1999)}]{Ezuka_and_Ishida1999}
{Ezuka}, H. \& {Ishida}, M. 1999, \apjs, 120, 277

\bibitem[{{Fabian} {et~al.}(1976){Fabian}, {Pringle}, \& {Rees}}]{Fabian1976}
{Fabian}, A.~C., {Pringle}, J.~E., \& {Rees}, M.~J. 1976, \mnras, 175, 43

\bibitem[{{George} \& {Fabian}(1991)}]{George_and_Fabian1991}
{George}, I.~M. \& {Fabian}, A.~C. 1991, \mnras, 249, 352

\bibitem[{{Girish} \& {Singh}(2012)}]{Girish_and_Singh1992}
{Girish}, V. \& {Singh}, K.~P. 2012, \mnras, 427, 458

\bibitem[{{Harrison} {et~al.}(2013){Harrison}, {Craig}, {Christensen}, {Hailey}, {Zhang}, {Boggs}, {Stern}, {Cook}, {Forster}, {Giommi}, {Grefenstette}, {Kim}, {Kitaguchi}, {Koglin}, {Madsen}, {Mao}, {Miyasaka}, {Mori}, {Perri}, {Pivovaroff}, {Puccetti}, {Rana}, {Westergaard}, {Willis}, {Zoglauer}, {An}, {Bachetti}, {Barri{\`e}re}, {Bellm}, {Bhalerao}, {Brejnholt}, {Fuerst}, {Liebe}, {Markwardt}, {Nynka}, {Vogel}, {Walton}, {Wik}, {Alexander}, {Cominsky}, {Hornschemeier}, {Hornstrup}, {Kaspi}, {Madejski}, {Matt}, {Molendi}, {Smith}, {Tomsick}, {Ajello}, {Ballantyne}, {Balokovi{\'c}}, {Barret}, {Bauer}, {Blandford}, {Brandt}, {Brenneman}, {Chiang}, {Chakrabarty}, {Chenevez}, {Comastri}, {Dufour}, {Elvis}, {Fabian}, {Farrah}, {Fryer}, {Gotthelf}, {Grindlay}, {Helfand}, {Krivonos}, {Meier}, {Miller}, {Natalucci}, {Ogle}, {Ofek}, {Ptak}, {Reynolds}, {Rigby}, {Tagliaferri}, {Thorsett}, {Treister}, \& {Urry}}]{Harrison_etal2013}
{Harrison}, F.~A., {Craig}, W.~W., {Christensen}, F.~E., {et~al.} 2013, \apj, 770, 103

\bibitem[{Hayashi {et~al.}(2011)Hayashi, Ishida, Terada, Bamba, \& Shionome}]{Hayashi_etal2011}
Hayashi, T., Ishida, M., Terada, Y., Bamba, A., \& Shionome, T. 2011, Publications of the Astronomical Society of Japan, 63, S739

\bibitem[{{Hayashi} {et~al.}(2018){Hayashi}, {Kitaguchi}, \& {Ishida}}]{Hayashi_etal2018}
{Hayashi}, T., {Kitaguchi}, T., \& {Ishida}, M. 2018, \mnras, 474, 1810

\bibitem[{{Hayashi} {et~al.}(2021){Hayashi}, {Kitaguchi}, \& {Ishida}}]{Hayashi_etal2021}
{Hayashi}, T., {Kitaguchi}, T., \& {Ishida}, M. 2021, \mnras, 504, 3651

\bibitem[{{Hellier}(1993)}]{Hellier1993}
{Hellier}, C. 1993, \mnras, 265, L35

\bibitem[{{HI4PI Collaboration} {et~al.}(2016){HI4PI Collaboration}, {Ben Bekhti}, {Fl{\"o}er}, {Keller}, {Kerp}, {Lenz}, {Winkel}, {Bailin}, {Calabretta}, {Dedes}, {Ford}, {Gibson}, {Haud}, {Janowiecki}, {Kalberla}, {Lockman}, {McClure-Griffiths}, {Murphy}, {Nakanishi}, {Pisano}, \& {Staveley-Smith}}]{HI4PICollaboration}
{HI4PI Collaboration}, {Ben Bekhti}, N., {Fl{\"o}er}, L., {et~al.} 2016, \aap, 594, A116

\bibitem[{{H{\={o}}shi}(1973)}]{Hoshi1973}
{H{\={o}}shi}, R. 1973, Progress of Theoretical Physics, 49, 776

\bibitem[{{Hua} \& {Lingenfelter}(1992)}]{Hua_and_Lingenfelter1992}
{Hua}, X.-M. \& {Lingenfelter}, R.~E. 1992, \apj, 397, 591

\bibitem[{{Islam} \& {Mukai}(2021)}]{Islam_and_Mukai2021}
{Islam}, N. \& {Mukai}, K. 2021, \apj, 919, 90

\bibitem[{{Iyer} {et~al.}(2015){Iyer}, {Mukherjee}, {Dewangan}, {Bhattacharya}, \& {Seetha}}]{Iyer2015}
{Iyer}, N., {Mukherjee}, D., {Dewangan}, G.~C., {Bhattacharya}, D., \& {Seetha}, S. 2015, \mnras, 454, 741

\bibitem[{{Jansen} {et~al.}(2001){Jansen}, {Lumb}, {Altieri}, {Clavel}, {Ehle}, {Erd}, {Gabriel}, {Guainazzi}, {Gondoin}, {Much}, {Munoz}, {Santos}, {Schartel}, {Texier}, \& {Vacanti}}]{Jansen_etal2001}
{Jansen}, F., {Lumb}, D., {Altieri}, B., {et~al.} 2001, \aap, 365, L1

\bibitem[{{Joshi} {et~al.}(2023){Joshi}, {Rawat}, {Schwope}, {Pandey}, {Scaringi}, {Sahu}, {Rao}, \& {Singh}}]{Joshi_etal2023}
{Joshi}, A., {Rawat}, N., {Schwope}, A., {et~al.} 2023, \mnras, 521, 6156

\bibitem[{{Kim} \& {Beuermann}(1995)}]{Kim1995}
{Kim}, Y. \& {Beuermann}, K. 1995, \aap, 298, 165

\bibitem[{{Lightman} \& {White}(1988)}]{Lightman_and_White1998}
{Lightman}, A.~P. \& {White}, T.~R. 1988, \apj, 335, 57

\bibitem[{{Lumb} {et~al.}(2012){Lumb}, {Schartel}, \& {Jansen}}]{Lumb_etal2012}
{Lumb}, D.~H., {Schartel}, N., \& {Jansen}, F.~A. 2012, arXiv e-prints, arXiv:1202.1651

\bibitem[{{Luna} {et~al.}(2018){Luna}, {Mukai}, {Orio}, \& {Zemko}}]{Luna_etal2018}
{Luna}, G.~J.~M., {Mukai}, K., {Orio}, M., \& {Zemko}, P. 2018, \apjl, 852, L8

\bibitem[{{Lutovinov} {et~al.}(2020){Lutovinov}, {Suleimanov}, {Manuel Luna}, {Sazonov}, {de Martino}, {Ducci}, {Doroshenko}, \& {Falanga}}]{2020Lutovinov}
{Lutovinov}, A., {Suleimanov}, V., {Manuel Luna}, G.~J., {et~al.} 2020, \nar, 91, 101547

\bibitem[{{Madsen} {et~al.}(2015){Madsen}, {Harrison}, {Markwardt}, {An}, {Grefenstette}, {Bachetti}, {Miyasaka}, {Kitaguchi}, {Bhalerao}, {Boggs}, {Christensen}, {Craig}, {Forster}, {Fuerst}, {Hailey}, {Perri}, {Puccetti}, {Rana}, {Stern}, {Walton}, {J{\o}rgen Westergaard}, \& {Zhang}}]{Madsen_etal2015}
{Madsen}, K.~K., {Harrison}, F.~A., {Markwardt}, C.~B., {et~al.} 2015, \apjs, 220, 8

\bibitem[{{Magdziarz} \& {Zdziarski}(1995)}]{Magdziarz_and_Zdziarski1995}
{Magdziarz}, P. \& {Zdziarski}, A.~A. 1995, \mnras, 273, 837

\bibitem[{{Maiolino} {et~al.}(2021){Maiolino}, {Titarchuk}, {Wang}, {Frontera}, \& {Orlandini}}]{Maiolino2021}
{Maiolino}, T., {Titarchuk}, L., {Wang}, W., {Frontera}, F., \& {Orlandini}, M. 2021, \apj, 911, 80

\bibitem[{{Masetti} {et~al.}(2006){Masetti}, {Morelli}, {Palazzi}, {Galaz}, {Bassani}, {Bazzano}, {Bird}, {Dean}, {Israel}, {Landi}, {Malizia}, {Minniti}, {Schiavone}, {Stephen}, {Ubertini}, \& {Walter}}]{Masetti2006}
{Masetti}, N., {Morelli}, L., {Palazzi}, E., {et~al.} 2006, \aap, 459, 21

\bibitem[{{Mukai}(2017)}]{Mukai2017}
{Mukai}, K. 2017, \pasp, 129, 062001

\bibitem[{{Mukai} {et~al.}(2015){Mukai}, {Rana}, {Bernardini}, \& {de Martino}}]{Mukai_etal_2015}
{Mukai}, K., {Rana}, V., {Bernardini}, F., \& {de Martino}, D. 2015, \apjl, 807, L30

\bibitem[{{Mushotzky} \& {Szymkowiak}(1988)}]{Mushotzky_and_Szymkowiak1988}
{Mushotzky}, R.~F. \& {Szymkowiak}, A.~E. 1988, in NATO Advanced Study Institute (ASI) Series C, Vol. 229, Cooling Flows in Clusters and Galaxies, ed. A.~C. {Fabian}, 53

\bibitem[{{Nauenberg}(1972)}]{Nauenberg1972}
{Nauenberg}, M. 1972, \apj, 175, 417

\bibitem[{{Norton} {et~al.}(1997){Norton}, {Hellier}, {Beardmore}, {Wheatley}, {Osborne}, \& {Taylor}}]{Norton1997}
{Norton}, A.~J., {Hellier}, C., {Beardmore}, A.~P., {et~al.} 1997, \mnras, 289, 362

\bibitem[{{Norton} \& {Watson}(1989)}]{Norton1989}
{Norton}, A.~J. \& {Watson}, M.~G. 1989, \mnras, 237, 853

\bibitem[{{Orlandini} {et~al.}(2012){Orlandini}, {Frontera}, {Masetti}, {Sguera}, \& {Sidoli}}]{Orlandini2012}
{Orlandini}, M., {Frontera}, F., {Masetti}, N., {Sguera}, V., \& {Sidoli}, L. 2012, \apj, 748, 86

\bibitem[{{Parker} {et~al.}(2005){Parker}, {Norton}, \& {Mukai}}]{Parker2005}
{Parker}, T.~L., {Norton}, A.~J., \& {Mukai}, K. 2005, \aap, 439, 213

\bibitem[{{Patterson}(1994)}]{Patterson1994}
{Patterson}, J. 1994, \pasp, 106, 209

\bibitem[{{Pek{\"o}n} \& {Balman}(2012)}]{Pekon2012}
{Pek{\"o}n}, Y. \& {Balman}, {\c{S}}. 2012, \aj, 144, 53

\bibitem[{{Pretorius}(2009)}]{Pretorius2009}
{Pretorius}, M.~L. 2009, \mnras, 395, 386

\bibitem[{{Reeves} {et~al.}(2008){Reeves}, {Done}, {Pounds}, {Terashima}, {Hayashida}, {Anabuki}, {Uchino}, \& {Turner}}]{Reeves2008}
{Reeves}, J., {Done}, C., {Pounds}, K., {et~al.} 2008, \mnras, 385, L108

\bibitem[{{Revnivtsev} {et~al.}(2004){Revnivtsev}, {Lutovinov}, {Suleimanov}, {Sunyaev}, \& {Zheleznyakov}}]{Revnivtsev_etal2004}
{Revnivtsev}, M., {Lutovinov}, A., {Suleimanov}, V., {Sunyaev}, R., \& {Zheleznyakov}, V. 2004, \aap, 426, 253

\bibitem[{{Rosen}(1992)}]{Rosen1992}
{Rosen}, S.~R. 1992, \mnras, 254, 493

\bibitem[{{Rothschild} {et~al.}(1981){Rothschild}, {Gruber}, {Knight}, {Matteson}, {Nolan}, {Swank}, {Holt}, {Serlemitsos}, {Mason}, \& {Tuohy}}]{Rothschild_etal1981}
{Rothschild}, R.~E., {Gruber}, D.~E., {Knight}, F.~K., {et~al.} 1981, \apj, 250, 723

\bibitem[{{Shaw} {et~al.}(2020){Shaw}, {Heinke}, {Mukai}, {Tomsick}, {Doroshenko}, {Suleimanov}, {Buisson}, {Gandhi}, {Grefenstette}, {Hare}, {Jiang}, {Ludlam}, {Rana}, \& {Sivakoff}}]{Shawn2020}
{Shaw}, A.~W., {Heinke}, C.~O., {Mukai}, K., {et~al.} 2020, \mnras, 498, 3457

\bibitem[{{Singh} {et~al.}(1996){Singh}, {White}, \& {Drake}}]{Singh1996}
{Singh}, K.~P., {White}, N.~E., \& {Drake}, S.~A. 1996, \apj, 456, 766

\bibitem[{{Suleimanov} {et~al.}(2019){Suleimanov}, {Doroshenko}, \& {Werner}}]{Suleimanov2019}
{Suleimanov}, V.~F., {Doroshenko}, V., \& {Werner}, K. 2019, \mnras, 482, 3622

\bibitem[{{Tomsick} {et~al.}(2006){Tomsick}, {Chaty}, {Rodriguez}, {Foschini}, {Walter}, \& {Kaaret}}]{Tomsick_etal2006}
{Tomsick}, J.~A., {Chaty}, S., {Rodriguez}, J., {et~al.} 2006, \apj, 647, 1309

\bibitem[{{van Teeseling} {et~al.}(1996){van Teeseling}, {Kaastra}, \& {Heise}}]{vanTeeseling_etal1996}
{van Teeseling}, A., {Kaastra}, J.~S., \& {Heise}, J. 1996, \aap, 312, 186

\bibitem[{{Warner}(1995)}]{Warner1995}
{Warner}, B. 1995, {Cataclysmic variable stars}, Vol.~28

\bibitem[{{Yuasa} {et~al.}(2010){Yuasa}, {Nakazawa}, {Makishima}, {Saitou}, {Ishida}, {Ebisawa}, {Mori}, \& {Yamada}}]{Yuasa_etal2010}
{Yuasa}, T., {Nakazawa}, K., {Makishima}, K., {et~al.} 2010, \aap, 520, A25

\end{thebibliography}

\end{document}